# Regge gravity on general triangulations

Wolfgang Beirl, Harald Markum, and Jürgen Riedler

*Institut für Kernphysik, Technische Universität Wien, A-1040 Vienna, Austria*

**Abstract**

We investigate quantum gravity in four dimensions using the Regge approach on triangulations of the four-torus with general, non-regular incidence matrices. We find that the simplicial lattice tends to develop spikes for vertices with low coordination numbers even for vanishing gravitational coupling. Different to the regular, hypercubic lattices almost exclusively used in previous studies, we find now that the observables depend on the measure. Computations with nonvanishing gravitational coupling still reveal the existence of a region with well-defined expectation values. However, the phase structure depends on the triangulation. Even with additional higher-order terms in the action the critical behavior of the system changes with varying (local) coordination numbers.

## 1 Introduction

The Regge calculus is an elegant method for investigations of quantum gravity through simplicial approximations [1, 2]. The sum over smooth geometries $\mathbf{g}$ in the Euclidean path integral

$$Z = \int D\mathbf{g} e^{-I_E(\mathbf{g})} \tag{1}$$

is replaced by a sum over simplicial lattices with the associated Regge action discretizing the gravitational action $I_E$ [3]. Simplicial lattices are specified by their incidence matrices, describing how the simplices are glued together, and the (squared) lengths of the contained links.

Two complementary approaches have been proposed to evaluate (numerically) simplicial path integrals varying either the incidence matrices for equal link lengths or, what is called Regge calculus, the link lengths for fixed triangulation. While the former method, known as dynamical triangulation, is most successful in two dimensions it might have serious problems in four dimensions [4]. The advantage of the Regge approach lies in the fact that the classical continuum limit is well-defined and quantization in the weak-field approximation can be studied [5].

Non-perturbative investigations of the Regge approach in four dimensions have been carried out on regular hypercubic triangulations of the 4-torus with rather astonishing results [6, 7, 8]:

(i) Despite the unbounded pure Regge action, a well-defined phase with finite expectation values exists for small gravitational coupling parameters [6, 7, 8].

(ii) Investigating a family of local measures, their influence on the well-defined phase was found to be small [8].

(iii) Including an additional higher-order $R^2$ term in the action, a phase transition of $2^{nd}$ order was found, indicating the existence of a non-trivial continuum limit [7].

(iv) The incorporation of additional matter fields does not affect the existence of the well-defined phase within the Regge approach [9].

Until now, the Regge calculus was applied almost exclusively for the regular hypercubic triangulation of the 4-torus. However, if one considers arbitrary topologies one has to use necessarily different triangulations and in general one would prefer to perform computations on random triangulations [6, 7]. If Regge gravity is a realis-tic theory the 'microscopic details' of the underlying triangulation should not affect physical expectation values.

Unfortunately, first studies of the Regge approach for non-regular incidence matrices signalled possible problems due to the occurence of spikes at vertices with low coordination number [10]. To clarify these findings, we investigate in this article simplicial quantum gravity on fixed but non-regular triangulations of the 4-torus, with the (squared) link lengths as dynamical variables. For this purpose we apply barycentric subdivision steps on a regularly triangulated 4-torus, inserting additional vertices with small local coordination number. We study the influence of the measure, the existence of the well-defined phase and the effect of additional $R^2$ terms.

## 2 Simplicial quantum gravity

To determine the *triangulation* of a given 4-manifold $\mathcal{M}$ one has to specify a set of $N_d$ d-simplices (i.e. vertices, links, triangles, tetrahedra and 4-simplices) which make up a simplicial manifold with the desired topology [3]. Until now, almost all numerical computations have been carried out on regular triangulations of the 4-torus, based on a hypercubic tesselation, with $N_0 \geq 3^4$, $N_1 = 15N_0$, $N_2 = 50N_0$, $N_3 = 60N_0$, and $N_4 = 24N_0$. Details of this construction and quantization in the weak-field approximation are discussed in the literature [5]. We call a triangulation *regular* if the local coordination numbers are the same for each vertex. In contrast to the hypercubic triangulation, we use in this study the construction of Kühnel and Lassmann, triangulating the 4-torus with $N_0 = 31$ vertices [11]. Most interestingly, the (global) coordination numbers (i.e. $N_1/N_0$ etc.) are the same as for the hypercubic case.

To construct a *non-regular* triangulation we have to change the local coordination numbers. We apply a number of barycentric subdivision (b.s.d.) steps, inserting additional vertices such that the total number of simplices increases as $\Delta N_0 = 1$, $\Delta N_1 = 5$, $\Delta N_2 = 10$, $\Delta N_3 = 10$, and $\Delta N_4 = 4$. While the number of links attached to a vertex is 30 for the regular triangulations it is 5 for the additional b.s.d. vertices [10].

Once the incidence matrix is fixed one can assign to each link $l$ the squared length $q_l$ to approximate a smooth 4-geometry. The infinitesimal line element $ds^2$ is thus replaced by the finite squared distances $\{q_l\}$ between neighboring vertices. We consider the Euclidean sector only, $q_l > 0$, and require that every 4-simplex can be embedded in $R^4$.

Given a Euclidean configuration $\{q_l\}$ several geometric quantities can be calculated. In the following we need the 4-volume $V_s$ of every 4-simplex $s$ and its fatness $\phi_s$ defined as

$$\phi_s = C^2 \frac{V_s{}^2}{\max_{l \in s}(q_l{}^4)}, \tag{2}$$

where we set the constant to $C = 24$. (Cheeger et al. use a slightly different but equivalent definition [12].) The fatness takes its maximum value of 0.3125 for equilateral simplices and tends to 0 for collapsing ones. Furthermore, we use the area $A_t$ of every triangle $t$ and the associated deficit angle $\delta_t$ given by

$$\delta_t = 2\pi - \sum_{s \supset t} \theta_{s,t}, \tag{3}$$

with the dihedral angle $\theta_{s,t}$ [2, 3, 7]. An important observable is the average curvature [7]

$$R = \frac{2 \sum_t A_t \delta_t}{\sum_s V_s} \frac{1}{N_1} \sum_l q_l \ . \tag{4}$$

The Einstein-Hilbert action is replaced by a sum over all triangles in the lattice

$$\int d^4x \sqrt{g} R \ \rightarrow \ 2 \sum_t A_t \delta_t. \tag{5}$$

Increasing the number of simplices and decreasing the link lengths such that the lattice skeleton approaches a smooth geometry, general relativity is recovered in the classical continuum limit if $\phi_s \geq f =$const$> 0$ [12, 13].

Concerning quantization, we assign concrete meaning to the functional integral (1) replacing the integration over closed 4-geometries by a sum over distinct simplicial lattices with fixed triangulation. To keep the 4-volume of the simplicial lattice finite one incorporates a cosmological-constant term in the action

$$- I_E(\{q_l\}) = \beta \sum_t A_t \delta_t - \lambda \sum_s V_s. \tag{6}$$

The variables $q_l$, $A_t$, $V_s$ are expressed in units of the bare Planck length, the parameter $\beta$ determines the gravitational coupling and $\lambda$ fixes the average lattice volume.

To complete the definition of the simplicial path integral one has to specify the integration measure. Within the Regge approach it seems natural to translate the gravitational measure into a product over the links of the simplicial lattice

$$Dq = \prod_l \mu(l) dq_l \mathcal{F}(q_1, ..., q_{N_1}), \tag{7}$$

where the function $\mathcal{F}$ is equal to 1 if the generalized triangle inequalities are fulfilled and 0 otherwise, ensuring that the computations are performed in the Euclidean sector [3, 6, 7]. The weight function $\mu$ is a (perhaps complicated and non-local) function of $\{q_l\}$ associated with each link $l$ [14, 15]. According to previous studies we consider as a local and simple ansatz the one-parameter family

$$\mu(l) = {q_l}^{\sigma-1}, \tag{8}$$

where $\sigma \geq 0$ determines the behavior of the measure under rescaling [8]. The case $\sigma = 1$ corresponds to the uniform measure and for $\sigma = 0$ a scale-invariant measure results.

Considering the behavior of the path integral

$$Z(\beta, \lambda) = \int \prod_l dq_l {q_l}^{\sigma-1} \mathcal{F}(\{q_l\}) e^{\beta \sum_t A_t \delta_t - \lambda \sum_s V_s} \tag{9}$$

under rescaling one obtains

$$-\beta \langle \sum_t A_t \delta_t \rangle + 2\lambda \langle \sum_s V_s \rangle = \sigma N_1 \tag{10}$$

in the well-defined phase [7]. We set the parameter $\lambda$ equal to $\sigma$ so that $\langle V_s \rangle$ takes the same value $N_1/(2N_4)$ at $\beta = 0$ for all $\sigma > 0$.

Numerical computations within the Regge approach are based on the Metropolis method generating a Markov chain of configurations $\{q_l\}_\tau$ according to the measure (8) and the action (6) [6, 7, 8]. After thermalization of the system, averages over the generated configurations approximate the expectation values. A priori one expects that simulations of simplicial quantum gravity with unbounded action might not find an equilibrium. However, several studies on regular hypercubic triangulations of the 4-torus show a well-defined phase with finite expectation values for $\beta < \beta_c$. This phase was found to be stable for different measures as long as $0 \lesssim \sigma \lesssim 1$ [7, 8].

## 3 Phase structure for non-regular triangulations

Previous studies indicate that vertices with low coordination number tend to develop spikes fluctuating with rather long auto-correlation lengths even for vanishing gravitational coupling [10]. To analyze the mechanism in more detail, we use a triangulation as small as possible allowing to perform 1M sweeps for each data point. We employ the construction of Kühnel and Lassmann [11] and apply three b.s.d. steps to neighboring simplices increasing the total number of vertices to $N_0 = 34$. We further use a lower limit for the fatness, $\phi_s \geq f > 0$, to restrict the configuration space.

Our first goal is to investigate the influence of the measure on the length of the spikes for vanishing gravitational coupling, $\beta = 0$. We performed computations for three different values of the measure parameter, $\sigma = 1, \frac{1}{2}, \frac{1}{10}$, and decreased the limit of the fatness stepwise by a factor $\frac{1}{2}$ as $f = 2^m 10^{-6}$, $m = 9, \ldots, 0$. To study explicitly the behavior of the inserted irregularities, we observed the expectation value $\langle q_i \rangle$ averaging the (squared) lengths of those links attached to the additionally inserted vertices. It measures directly the next-neighbor distances and thus the lengths of the developing spikes. This quantity can be compared with the expectation value $\langle q_r \rangle$ averaging the (squared) lengths of all other links that do not touch the additional vertices. For all three considered measures this expectation value stays rather constant, $\langle q_r \rangle \sim 5.1$, independent of the cutoff $f$. This was also found for the completely regular triangulation.

Fig. 1 shows $\langle q_i \rangle$ as a function of $f$ and one realizes that the heigth of the spikes increases with decreasing $f$. By means of numerical methods one cannot decide whether $\langle q_i \rangle$ diverges in the limit $f \to 0$ or not. However, $\langle q_i \rangle$ decreases with smaller $\sigma$ and the results show clearly that in the scale invariant limit, $\sigma \to 0$, the tendency to develop spikes at the inserted irregularities is suppressed. Thus, we use the 'nearly scale invariant' measure, $\sigma = \frac{1}{10}$, also for computations with non-vanishing gravitational coupling and apply the lowest limit on the fatness, $f = 10^{-6}$.

Fig. 2a displays the average link length $\langle q_i \rangle$ at the inserted vertices as a function of $\beta$ in comparison to the expectation value $\langle q_r \rangle$ of the regular part. At $\beta > 0$ for every data point 1.1M sweeps have been performed and after 100k thermalization steps every $100^{th}$ sweep has been recorded and used for the average values. A rather interesting behavior is observed: As long as $\beta < \beta_1 \sim 0.025$ a phase is found with monotonically increasing values of $\langle q_i \rangle$ while $\langle q_r \rangle$ stays almost constant (in fact it slightly decreases for $\beta < 0.02$). At $\beta_1$ both expectation values suddenly jump to significantly larger values and enter between $\beta_1 < \beta < \beta_2 \sim 0.04$ a region with $\langle q_i \rangle \sim 10^3$ and $\langle q_r \rangle \sim 10$. At $\beta_2$ the system performs a second jump into an ill-defined phase with diverging observables, known from the computations on regular hypercubic triangulations, which one can resolve for a finite cut off $f > 0$ only.

However, since the rotation to the Euclidean sector is not uniquely defined for quantum gravity, the sign of the coupling parameter $\beta$ is not fixed a priori [3, 6]. We therefore present results also for negative gravitational couplings, $\beta < 0$, in Fig. 2a. Decreasing $\beta$ stepwise, one observes a well-defined phase with homogeneous configurations, $\langle q_i \rangle \sim \langle q_r \rangle$. Thus, only 200k sweeps after 50k thermalization steps have been performed and measurement has been taken every $10^{th}$ sweep for each data point. At $\beta_0 \sim -0.11$ a transition to an ill-defined region with collapsing simplices occurs. Since the lattice size is small, it does not make sense to try to determine the order of this phase transition and we postpone this question to further studies.

To obtain further insight we plot the expectation value $\langle \phi_s \rangle$ of the fatness as a function of $\beta$. This observable (3) does not distinguish between 4-simplices containing irregular vertices. Nevertheless, all three transitions are reflected as seen in Fig. 2b.

An important physical quantity is the expectation value $\langle R \rangle$ of the curvature as a function of $\beta$. In Fig. 2c we observe an increase at $\beta_1 \sim 0.025$ where the link lengths at the extra vertices become rather large. It is followed by another increase at $\beta_2 \sim 0.04$ where the remaining regular vertices tend to develop spikes. This second transition represents the well-known transition from the well-defined to the ill-defined phase of the regularly triangulated Regge theory on hypercubes. Turning to negative gravitational couplings we find a decrease of $\langle R \rangle$ at $\beta_0 \sim -0.11$ to an ill-defined phase with growth of the link lengths at both types of vertices.

## 4 Higher-order terms for non-regular Regge lattices

The introduction of an additional $R^2$ term was initially proposed to cure the problem of the unbounded Regge action. However, there is numerical evidence that a phase with well-defined finite expectation values exists for simplicial quantum gravity even without such a term. Moreover, it has been claimed that an additional higher-order term ensures that the transition from the well-defined phase to the region of 'rough geometries' is of $2^{nd}$ order, accompanied by large correlation lengths, indicating a non-trivial continuum limit [7].

At this point it is important to remember that the Regge approach deals with a large number of unphysical degrees of freedom [5, 16]. On the triangulated hypercubic 4-torus one has 15 link lengths per vertex in contrast to the 10 independent components of the metric tensor. Notice that 4 of these 10 components can be fixed by choice of coordinates and the remaining 6 degrees of freedom are still subject of constraints reflecting the Bianchi identities. Respecting coordinate invariance, i.e. the diffeomorphism group, within the Regge calculus would lead to a non-trivial and non-local measure [15]. But until now, numerical studies within the Regge approach have ignored these difficulties and this might be justified far from the continuum limit with the diffeomorphism symmetry completely broken. However, if one reaches the continuum limit, this question becomes important. (For a description of these problems in two dimensions see Refs. [16, 17].) Therefore, it is interesting to study higher-order gravity on general, non-regular triangulations with varying coordination numbers and degrees of freedom. If a continuum limit can be reached, recovering the diffeomorphism group, the universal behavior of the Regge approach should not depend on the choice of the incidence matrix.

We thus now study the influence of the lattice construction in the presence of an additional term $-a\sum_t A_t^2\delta_t^2 V_t^{-1}$ in the action with the 4-volume per triangle $V_t$ given by barycentric subdivision [7]. Following the literature we set $a = 0.005$ where a $1^{st}$ order transition at $\langle A_t\delta_t\rangle \sim 0$ was ruled out in favor of a $2^{nd}$ order transition [7]. To allow for a comparison with the results of hypercubic triangulation we use the uniform measure, $\sigma = \lambda = 1$. Our computation completely parallels those of Ref. [7] with the minor exception of posing a lower limit on the fatness, $\phi_s > f = 10^{-5}$, and the major difference of using triangulations with $N_0 = 31, 34$ and $40$ vertices, i.e.

the regular and two non-regular triangulations with 3 and 9 additional vertices.

In Fig. 3 we present results for the curvature on the regularly triangulated lattice, $N_0 = 31$, which seems to behave as described in the literature [7]. At $\beta \sim 0.25$ the system reaches $\langle R \rangle \sim 0$ and a comparison of our data shows good agreement in the well-defined region, $\langle R \rangle < 0$, $\beta < 0.25$. Different to the literature, we found stable expectation values even for $\langle R \rangle > 0$ and would locate the phase transition at $\beta > 0.28$. However, this is of minor significance since the lattice size is small, but indicates that it will be difficult to determine the order of the phase transition even for regular triangulations.

Considering on the other hand the non-regular triangulations one observes that the additional action term suppresses the spikes at the inserted vertices. Fig. 4a compares directly the expectation value $\langle q_i \rangle$ of the link lengths at the irregular vertices with $\langle q_r \rangle$ averaging over the links of the regular part for $N_0 = 34$. Surprisingly, $\langle q_i \rangle$ stays below $\langle q_r \rangle$ for small $\beta$, and we reduced in all $R^2$-term simulations the number of iterations per data point performing 50k sweeps to reach equilibrium and 200k sweeps to gather statistics while measuring expectation values every $10^{th}$ step. As in the case without $R^2$ terms, first the extra vertices tend to build spikes at $\beta_1 \sim 0.16$ followed by the remaining vertices at $\beta_2 \sim 0.25$.

The expectation value $\langle \phi_s \rangle$ averaged over the whole lattice does not distinguish between the different types of vertices, but reflects the behavior of $\langle q_i \rangle$ and $\langle q_r \rangle$. Fig. 4b suggests a transition at $\beta_1 \sim 0.16$ from the vertices with low coordination number and indicates a second transition to occur at $\beta_2 \sim 0.25$ with all 4-simplices involved.

The 3 additional vertices lead to a shift in the average curvature towards positive values and as seen in Fig. 4c $\langle R \rangle$ jumps at $\beta_1 \sim 0.16$ from $\langle R \rangle \sim 2.6$ to $\langle R \rangle \sim 40.3$. A preliminary computation with 9 inserted vertices, $N_0 = 40$, exhibits a further shift towards larger curvature. A comparison of Fig. 4c with 4a shows that the transition at $\beta_1$ affects only the additional vertices whereas the transition at $\beta_2$ from the regular vertices leads to an increase of $\langle R \rangle$ that is unlimited in principle for $f \to 0$.

To summarize, the effect of vertices inserted via b.s.d. steps is similar as in the pure Regge action discussed in the previous section. The additional higher-order term only shifts the critical values $\beta_1, \beta_2$. Since the lattice size is small in our study, one has to be aware of the preliminary character of our findings. However, the results seem to be of importance for future simulations within the Regge approach for random triangulations on larger lattices. In general, for non-regular triangulations with varying local coordination numbers the influence of the incidence matrix on the phase transition seems to be essential: Even with an additional $R^2$ term the possible existence of a $2^{nd}$ order phase transition at $R \sim 0$, as observed for regular triangulations, is covered by another transition that occurs at vertices with low coordination number.

## 5 Conclusion

The Regge calculus provides a direct route to systematic approximations of the quantum-gravity path-integral. In a general setting one would investigate the path integral on random triangulations with arbitrary topologies. As a first step in this direction we considered non-regular triangulations by inserting additional vertices into a regularly triangulated 4-torus using barycentric subdivisions.

We find that for vanishing gravitational coupling the inserted vertices develop spikes, as observed in earlier studies. Different to the regular, hypercubic triangulation, the influence of the measure becomes non-negligible, such that the formation of spikes is suppressed in the scale invariant limit.

Using a nearly scale invariant measure we observe for negative gravitational couplings, $0 > \beta > \beta_0 \sim -0.1$, a well-defined phase characterized by homogeneous configurations with the spikes at the irregularly inserted vertices strongly damped. If this property holds for arbitrary random triangulations one should investigate this phase more detailed, especially the character of the transition at $\beta_0$. For positive couplings we observe two transitions, at $\beta_1 \sim 0.025$ and $\beta_2 \sim 0.04$, indicated by sudden changes of the average link lengths at the extra and regular vertices, respectively. The spikes at the vertices with low local coordination numbers are significantly larger than the average link length.

The introduction of an additional higher-order $R^2$ term in the action with $a = 0.005$ strongly suppresses these spikes. However, the critical behavior of the system seems to differ from that observed on regular, hypercubic triangulations. As for pure gravity, the additional irregularities undergo a transition, at $\beta_1 \sim 0.16$, in addition to that observed for the regularly triangulated part at $\beta_2 \sim 0.25$. In future work, the phase structure of the Regge theory has to be examined for general triangulations and two main questions have to be dealt with: Do random lattices exhibit several transitions? Do the (two or more) transitions merge on larger lattices?

## Acknowledgment

This work was supported in part by "Fonds zur Förderung der wissenschaftlichen Forschung" under Contract P9522-PHY. Part of the computations have been performed at the workstation cluster in SCRI at FSU, Tallahassee. W. Beirl thanks B. Berg and W. Bock for helpful and encouraging discussions.

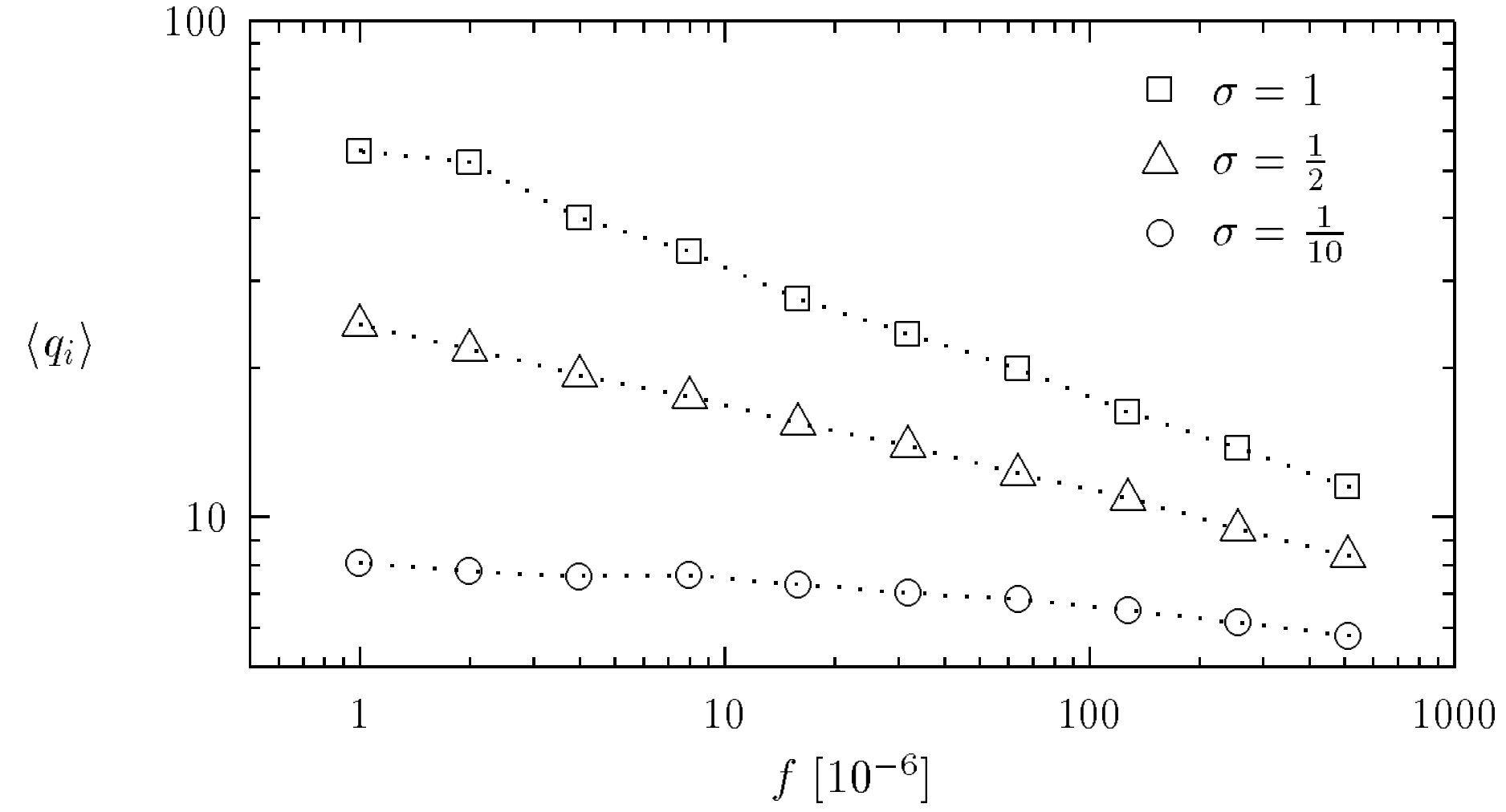

Fig. 1. Dependence of the simulation on the measure for vanishing gravitational coupling. A non-regular triangulation of the 4-torus based on the construction of Kühnel and Lassmann [11] with 3 additional vertices obtained by barycentric subdivision has been considered. For each value of the measure parameter $\sigma = 1, \frac{1}{2}$, and $\frac{1}{10}$ the lower limit of the fatness is decreased stepwise by a factor $\frac{1}{2}$ towards $10^{-6}$. The average next-neighbor distance of the additional vertices diverges for $f \to 0$ as long as $\sigma$ is significantly larger than 0. Notice the logarithmic scale. Statistical error bars are in the size of the symbols.

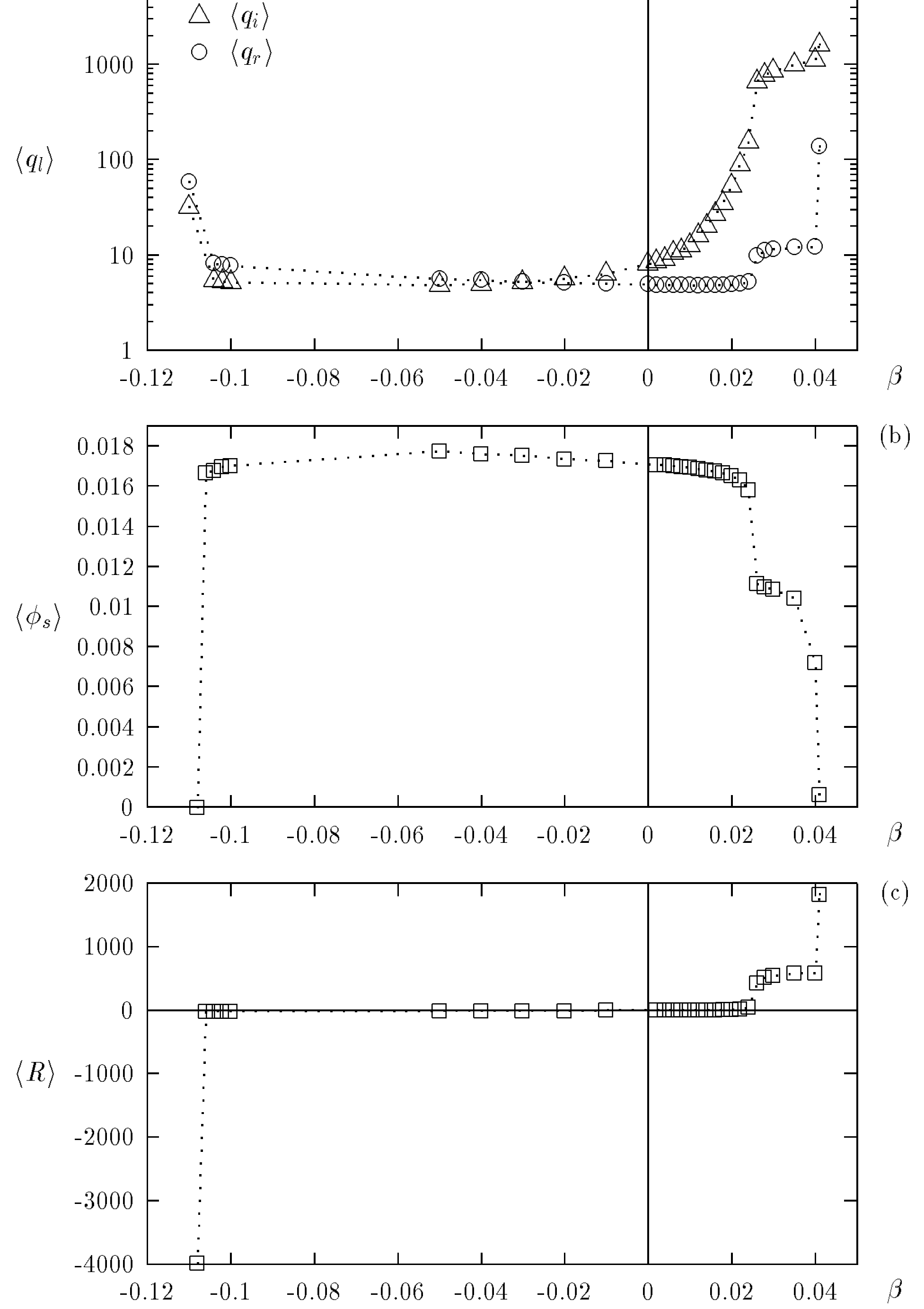

Fig. 2. Simulation for nonvanishing gravitational coupling $\beta$. The average squared link length $\langle q_i \rangle$ at the additional vertices and $\langle q_r \rangle$ of the regularly triangulated part are shown in plot (a) in a logarithmic scale. The system undergoes transitions in the positive coupling region at $\beta_1 \sim 0.025$ where the extra vertices develop spikes and at $\beta_2 \sim 0.04$ where the rest follows. For negative couplings the well-defined phase continues with homogeneous configurations, $\langle q_i \rangle \sim \langle q_r \rangle$, with the spikes at the additional vertices suppressed. At $\beta_0 \sim -0.11$ a transition to an ill-defined region with collapsing simplices is observed. The expectation values of the fatness $\langle \phi_s \rangle$ (b) and of the curvature $\langle R \rangle$ (c) are accordingly influenced at $\beta_0, \beta_1$, and $\beta_2$.

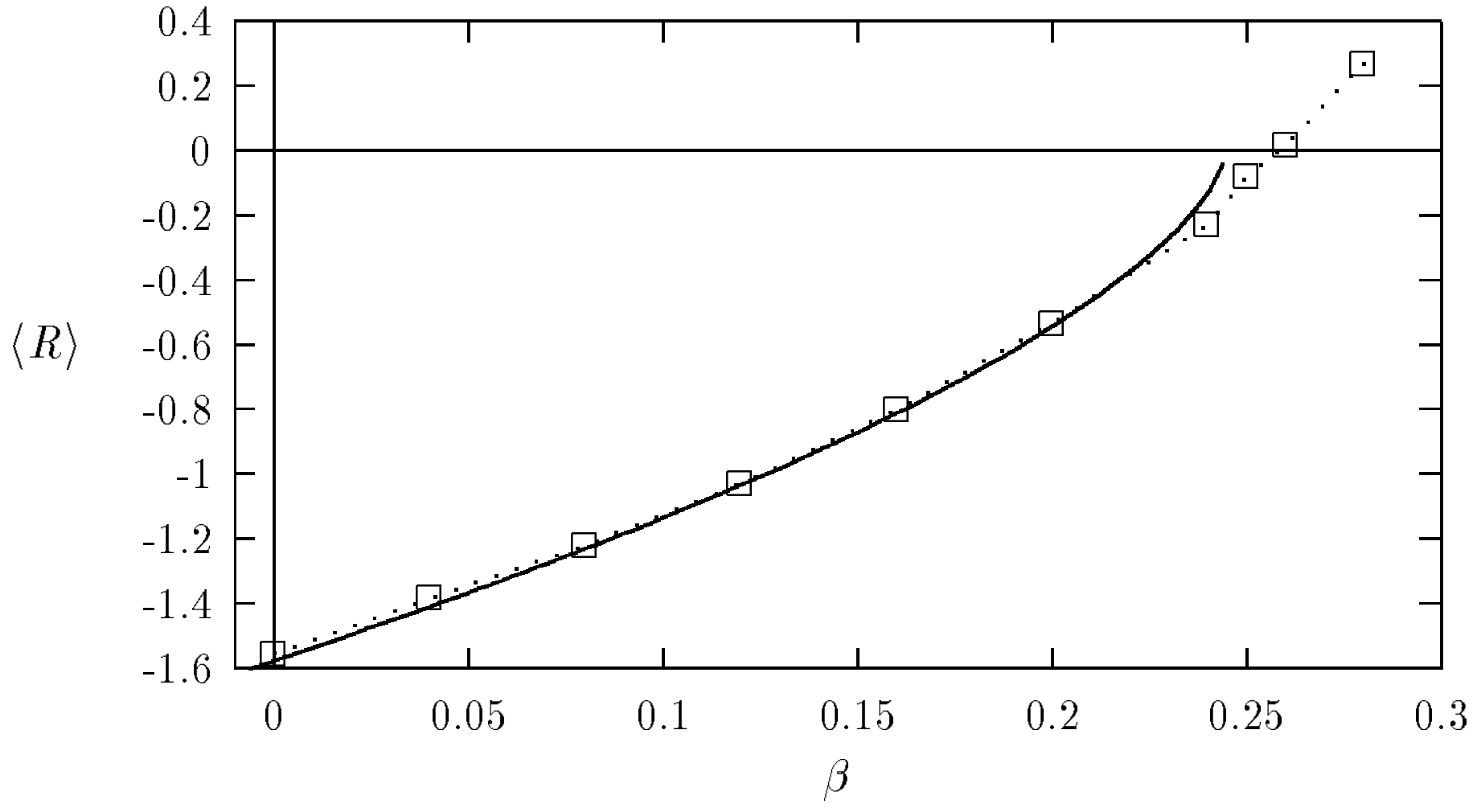

Fig. 3. Simulation with additional $R^2$ term. The scale invariant curvature $\langle R \rangle$ is displayed as a function of $\beta$. The system reaches $\langle R \rangle \sim 0$ at $\beta_2 \sim 0.25$. Our data is compared to $R(\beta) = -A(\beta_c - \beta)^\delta$ with $A = 3.794$, $\beta_c = 0.2443$, $\delta = 0.624$ as reported in [7].

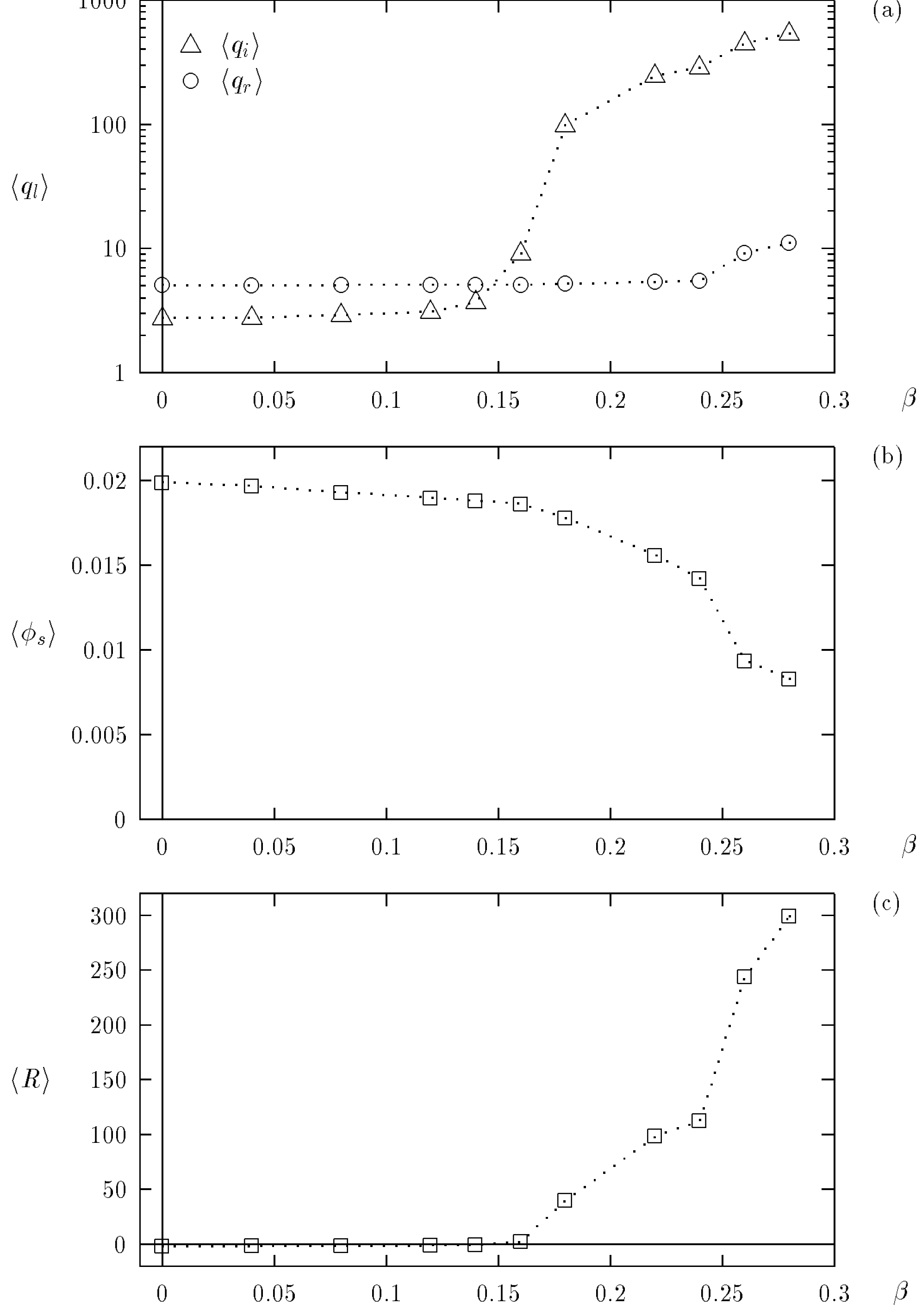

Fig. 4. Simulation with additional $R^2$ term. The average squared link length $\langle q_i \rangle$ at the additional vertices and $\langle q_r \rangle$ of the regularly triangulated part are shown in plot (a) in a logarithmic scale. At $\beta_1 \sim 0.16$ a transition of the link lengths at the additional vertices is observed and at $\beta_2 \sim 0.25$ the remaining links start to grow. The expectation values of the fatness $\langle \phi_s \rangle$ (b) and of the curvature $\langle R \rangle$ (c) are also influenced at $\beta_1$ and $\beta_2$.